\documentclass[aip,jcp,preprint,superscriptaddress,amsmath,amssymb]{revtex4-2}
\usepackage{graphicx}% Include figure files
\usepackage{xcolor}
\usepackage{dcolumn}% Align table columns on decimal point
\usepackage{bm}% bold math
\usepackage[utf8]{inputenc}
\usepackage[T1]{fontenc}
\usepackage{etoolbox}
\usepackage[bookmarksopen=true,bookmarksnumbered=true,colorlinks=true,urlcolor=blue,linkcolor=blue,citecolor=blue]{hyperref}% add hypertext capabilities
\usepackage{cleveref}

\makeatletter
\def\@email#1#2{%
 \endgroup
 \patchcmd{\titleblock@produce}
  {\frontmatter@RRAPformat}
  {\frontmatter@RRAPformat{\produce@RRAP{*#1\href{mailto:#2}{#2}}}\frontmatter@RRAPformat}
  {}{}
}%
\makeatother

\begin{document}
\title{Quantum hot carrier spectra in plasmonic catalysis}
% Force line breaks with \\
%\thanks{A footnote to the article title}%
\author{Yu Chen}
\altaffiliation{These authors contributed equally}		
\author{Hanwen Jin}
\altaffiliation{These authors contributed equally}
\affiliation{Beijing Computational Science Research Center, Beijing 100193, China}
\author{Fei Gao}
\affiliation{Center of Single-Molecule Sciences, Institute of Modern Optics, Frontiers Science Center for New Organic Matter, Tianjin Key Laboratory of Micro-Scale Optical Information Science and Technology, College of Electronic Information and Optical Engineering, Nankai University, Tianjin 300350, China}
\affiliation{Donostia International Physics Center, Manuel Lardizabal Ibilbidea 4,Donostia-San Sebastián 20018, Spain}
\author{Johannes Lischner}
\affiliation{Department of Materials and the Thomas Young Centre for Theory and Simulation of Materials, Imperial College London, London SW7 2AZ, United Kingdom}
\author{Shiwu Gao}
\email{swgao@csrc.ac.cn}
\affiliation{Beijing Computational Science Research Center, Beijing 100193, China}
\date{\today}% It is always \today, today,
% but any date may be explicitly specified

\begin{abstract}
	Vibrational activation of admolecules on metal nanoparticles is an elementary step in plasmonic catalysis, yet the underlying dynamics driven by hot carriers is not fully understood in the quantum regime. Using an atomistic description of plasmonic hot carrier generation, we investigate vibrational excitation and dissociation of oxygen on silver nanoparticles as a function of diameter $D$. As $D$ reduces from the classical to quantum-sized regime, quantized distribution of hot carriers emerges with increasing population in the high-energy regions. These highly energetic hot carriers deliver more efficient vibrational coupling and dissociation. The rate of vibrational excitation shows a linear $1/D$ scaling, which results from Landau damping. It turns nonlinear at elevated light intensities due to vibrational heating generated by multiple electron scattering. The finding of quantized distribution of hot carrier in plasmonic catalysis opens new avenues for selective control and nonthermal energy conversion.
\end{abstract}
\maketitle

\section{\label{sec:Intro} Introduction}
Quantum confinement in small metal nanoparticles (NPs) leads to quantization in electronic energy spectra. The large surface-to-volume ratio supports abundant surface states, which provide strong surface coupling, sensitivity, and reactivity. Single-atom catalyst represents the extreme limit of such strong coupling, which could dramatically enhance catalytic reactions by modifying the interatomic bonding and reducing the reaction barriers.~\cite{CH_2020,CH_2025} Dynamically, strong surface coupling also facilitates electron-molecule scattering and nonadiabatic energy transfer.~\cite{Nordlander2024} This work explores the electronic activation of molecular vibration and bond dissociation using the nonequilibrium hot electron-hole pairs generated by the Landau damping of surface plasmons. These energetic hot carriers (HCs) are under intensive ongoing exploration, and are promising for vibrational energy transfer in adsorbed molecules~\cite{Kim2026,Linic2016_1,Dong2024,Kim2023,Kim2024} and photocatalytic reactions.~\cite{CF_2020,CO2_2025,Baldi2025,NH3_2018,NH3_2022,NH3_2024,NH3_2026} We propose and demonstrate that the NPs approaching the quantum-size regime provide a uniquely promising route for efficient and selective control of vibrational excitation thanks to the quantum nature of the HC distributions.

Nonequilibrium HC distributions can be created by continuous pumping and relaxation of the surface plasmon resonances under light illumination. Such HCs have been observed in recent experiments~\cite{Nth_exp_2025,Nth_exp_2020,Nth_exp_2022,Nth_exp_2018} and theoretically predicted in various nanostructures.~\cite{Seideman2013,Govorov2013,Manjavacas2014,Govorov2017,Govorov2020,Narang2014,Louie2015,Wang2015,Erhart2020,Lischner2022,Lischner2023_1} At interfaces, the high-energy HCs can be efficiently injected into adsorbed molecules~\cite{Link2019,Yang2024,Cortes2025,Erhart2019_1,Schatz2025} and semiconductor substrates~\cite{Lian2015,Atwater2020,Friedrich2025,Lian2021,Lian2025,Gao2019,Erhart2019_2} prior to relaxation. Intuitively, HC generation and interface scattering should become more efficient in smaller NPs due to the suppressed radiation damping,~\cite{Hartland2019} quantized electronic spectra,~\cite{PR2012,PR2018,PR2024} and the reduced phase spaces for electron-electron and electron-phonon scattering.~\cite{Link1999,Broyer2000,Treguer2003} These combined effects shall increase the lifetime of the energetic HCs and favor a more nonequilibrium distribution.~\cite{Nth_exp_2025,Schatz2026,Abajo2016,Liu2017} How such HCs evolve from the classical to the quantum-sized regime, and how this evolution alters the plasmonic reactions, is still an open question.

The Anderson-Newns model~\cite{Newns1969} provides a quantitative framework for electron-molecule scattering,~\cite{Nordlander2024} and nonadiabatic charge and energy transfers.~\cite{Gadzuk1991,Gao1997,Olsen2009,Christopher2014,Alabastri2024} Recently, we have extended this model to describe plasmonic catalysis using a unified description of thermal and nonthermal HC distributions.~\cite{Wu2022,Gao2025} In this work, we adopt a tight-binding description of electronic structures and HC generations in NPs.~\cite{Lischner2022,Lischner2023_1,Lischner2023_2,Lischner2024,Lischner2026} This atomistic approach incorporates the interband transitions and lifts the unphysical energy degeneracy of the electron gas model.~\cite{Govorov2013,Manjavacas2014,Govorov2017,Govorov2020} With the kernel-polynomial expression of HC generation,~\cite{Lischner2022,Lischner2023_1} the atomistic model has enabled us to treat large NPs.

By investigating the evolution of HC distribution as a function of particle size, the quantum features of HC spectra have been revealed in the small-size regime. These quantum HC spectra significantly enhance vibrational coupling and photodissociation, as demonstrated here for O$_2$ on Ag NPs as a prototypical system. Under low-light-intensity conditions, the photodissociation efficiency follows a linear $1/D$ scaling, attributable to the size-dependent Landau damping of surface plasmons.~\cite{Govorov2017,Govorov2020} Under intense illumination, however, this scaling turns nonlinear, driven by multiple electron-molecule scatterings and the population of vibrationally excited states. In the small-size limit ($D<5.0$~nm), the highly nonequilibrium HC energy distribution exhibits sensitive coupling to molecular resonances, leading to quantum oscillations in reaction rates. The impact of quantum HC distributions on photodissociation opens promising avenues for the selective control of vibrational excitation and plasmonic energy conversion.

\section{\label{sec:Method} Methods}
Our formulation in the electron gas model has been published previously.~\cite{Wu2022,Gao2025} Here, we extend it to atomistic model with tight-binding description of metal NPs. A nonthermal HC distribution $f_\mathrm{nth}(\epsilon)$ is introduced to describe the density response $\delta\rho(\epsilon)$ at energy $\epsilon$ under plasmon excitation.~\cite{Govorov2020,Wu2022,Gao2025} To enable effective computation of HC generation, the kernel polynomial method (KPM)~\cite{RMP2006} has been used to evaluate $\delta\rho(\epsilon)$,~\cite{Lischner2022,Lischner2023_1,Lischner2023_2,Lischner2024,Lischner2026} which avoids full diagonalization of the Hamiltonian. This scheme is efficient to treat large NPs with millions of atoms,~\cite{Lischner2022,Lischner2023_1} allowing us to obtain the HC spectra across a wide range of diameters. In KPM, the density of states $g(\epsilon)$ is expressed as
\begin{equation}\label{eqn:DOS_KPM}
	g(\epsilon)=\mathrm{Tr}[\delta(\epsilon-H_\mathrm{NP})],
\end{equation}
where $H_\mathrm{NP}$ is the tight-binding Hamiltonian. The expression of $\delta\rho(\epsilon)$ is recast into a KPM-compatible form as
\begin{equation}\label{eqn:delta_rho_KPM}
	\begin{split}
		\delta\rho(\epsilon)&=\frac{4\tau_e}{\tau_p}\int\mathrm{d}\mathcal{E}\int\mathrm{d}\mathcal{E}^\prime[f_0(\mathcal{E}^\prime)-f_0(\mathcal{E})]\phi(\mathcal{E},\mathcal{E}^\prime,\omega)\delta(\epsilon-\mathcal{E})\\& \hspace{3em}\times\left[\frac{1}{(\hbar\omega-\mathcal{E}+\mathcal{E}^\prime)^2+(\hbar/\tau_p)^2}+\frac{1}{(\hbar\omega+\mathcal{E}-\mathcal{E}^\prime)^2+(\hbar/\tau_p)^2}\right]
	\end{split}.
\end{equation}
Here, $\tau_e$ and $\tau_p$ denote the energy- and momentum-relaxation lifetimes of HCs, respectively,~\cite{Govorov2017} $f_0$ is the Fermi-Dirac distribution function of the ground state, and $\phi(\mathcal{E},\mathcal{E}^\prime,\omega)$ represents the energy-resolved optical matrix element at photon frequency $\omega$. The details of the numerical implementation are provided in the supplementary material.

The HC-induced inelastic transition rates for vibrational excitation and dissociation is given by~\cite{Gadzuk1991,Gao1997}
\begin{equation}\label{eqn:vib_exc}
	\begin{split}
		W_{n\to n^\prime}&=\frac{4\Delta_a^2}{\pi\hbar}\int\mathrm{d}\epsilon\,f(\epsilon)\left[1-f(\epsilon+(n-n^\prime)\hbar\Omega)\right]\\
		&\hspace{4em}\times\left|\sum_m\frac{\langle n^\prime|m\rangle\langle m|n\rangle}{\epsilon+(n-m)\hbar\Omega-\epsilon_a+i\Delta_a}\right|^2
	\end{split},
\end{equation}
where $\langle n^\prime|m\rangle\langle m|n\rangle$ denotes the Franck-Condon overlap, with $|n\rangle$ ($|n^\prime\rangle$) and $|m\rangle$ the vibrational states in the ground and excited potential energy surfaces, respectively. We have included both the thermal and nonthermal distribution functions on equal footing in Eq.~\eqref{eqn:vib_exc} as $f(\epsilon)=f_0(\epsilon)+f_\mathrm{nth}(\epsilon)$, and $f_\mathrm{nth}(\epsilon)=\delta\rho(\epsilon)/g(\epsilon)$.~\cite{Govorov2020,Wu2022,Gao2025} To model O$_2$ on Ag NPs, we used basically the following parameters:~\cite{O2_2011,O2_2012,Govorov2017} the vibrational quanta $\hbar\Omega$=0.10~eV, the molecular resonance energy $\epsilon_a$=2.40~eV and resonance broadening $\Delta_a$=0.60~eV, $\tau_e$=500~fs, $\tau_p$=33~fs, and an electron-vibration coupling constant of 0.05~eV. In addition, all results are checked with variable parameters to explore possible variations in NPs. 

\section{\label{sec:RD} Results}
\subsection{\label{sec:RD1} Evolution of hot carrier energy distribution}
\begin{figure*}[h!tbp]
  \includegraphics[width=\columnwidth]{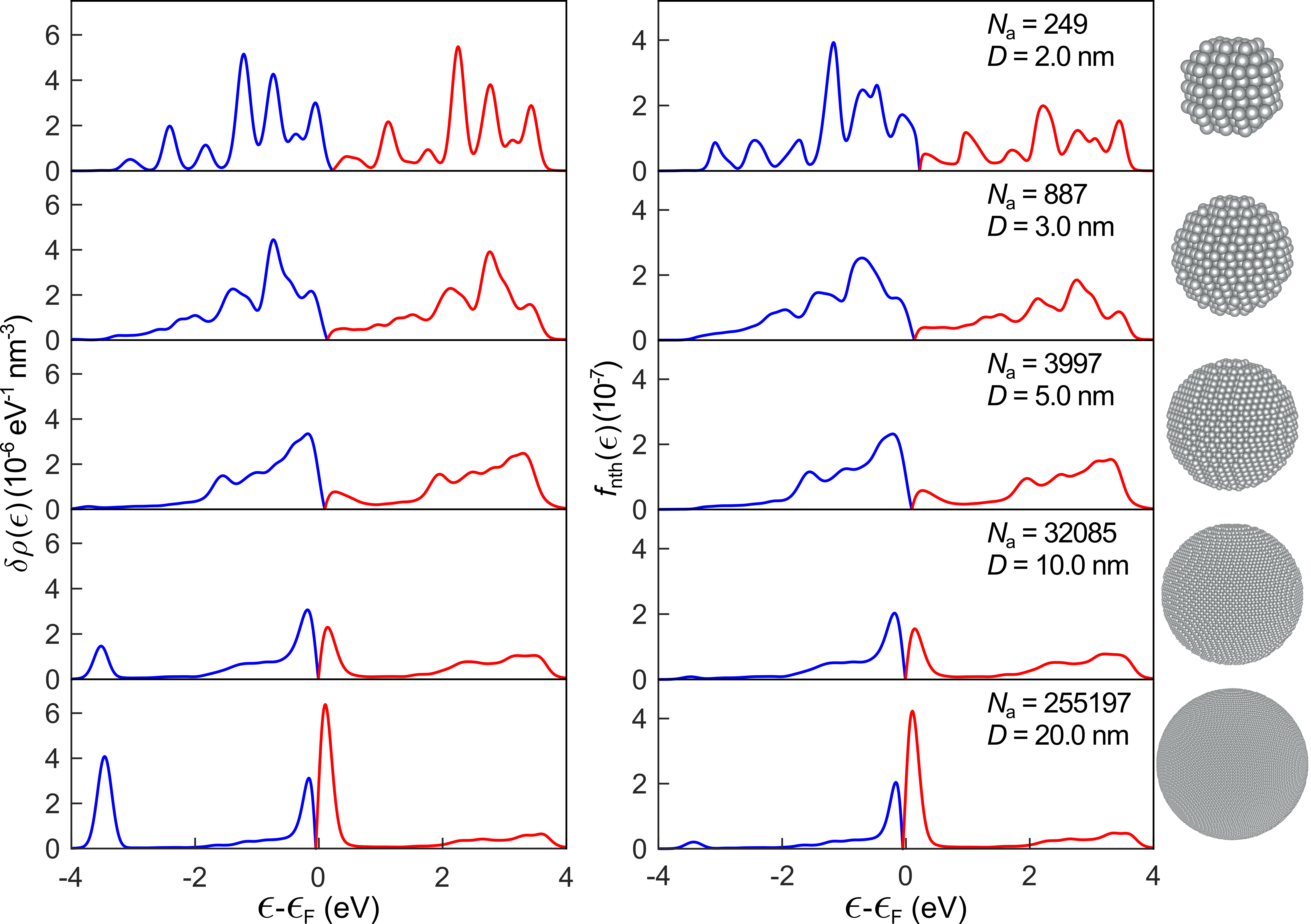}
	\caption{\label{fig:fig1} The density response $\delta\rho(\epsilon)$ and nonthermal distribution function $f_\mathrm{nth}(\epsilon)$ of plasmonic HCs evaluated in atomistic model. Five representative NPs with given diameters $D$ and atom numbers $N_a$ are presented. Their corresponding atomic structures are shown alongside. Hot electrons and holes are indicated by red and blue curves, respectively, with the hole distribution presented in opposite sign. All spectra are obtained with an intensity of 10$^6$~W/m$^2$.}
\end{figure*}

Figure~\ref{fig:fig1} displays the evolution of HC distributions as a function of $D$ ranging from 20.0~nm to 2.0~nm, which corresponds to classical and quantum-sized regimes. The right column shows the atomic structures used in the tight-binding calculations with parameters taken from the literature.~\cite{Papaconstantopoulos2015} Under resonant excitation of plasmon frequency $\hbar\omega_\mathrm{p}=3.5$~eV, the electrons are excited from [$-\hbar\omega_\mathrm{p}$, 0] to [0, $\hbar\omega_\mathrm{p}$] relative to the Fermi level. At $D=20.0$~nm, the HC distribution is characterized by mostly a smooth distribution with two peaks near the Fermi level. The smooth distribution comes from the intraband transitions.~\cite{Govorov2013,Manjavacas2014,Govorov2017,Govorov2020} In addition, there is an additional hole peak at -3.5~eV resulting from the interband transition from the d-bands to the Fermi level.~\cite{Wang2015,Erhart2020,Lischner2022,Lischner2023_1,Narang2014,Louie2015} Larger NPs have also been calculated. The features of HC spectra do not change appreciably beyond $D=20.0$~nm.~\cite{Lischner2022} As $D$ decreases, the HC spectrum evolves from a continuous distribution to more discrete peaks due to electron energy quantization in reduced dimension (Fig.~S1). Simultaneously, the sharp peaks near the Fermi level gradually diminish and then disappear at $D=5.0$~nm. The reduction and disappearance result from the narrowing of the d-band width and the downward shift of the d-band edge (Fig.~S2),~\cite{Visikovskiy2011} which prohibits the interband transitions contributing to the hot electron generation. These characteristics are a direct consequence of reduced interaction and electron energy quantization in small NPs. As another general feature, the fraction of hot electrons at high-energy states increases as $D$ decreases.~\cite{Manjavacas2014,Govorov2017,Govorov2020} For example, $f_\mathrm{nth}(\epsilon)$ at 3.5~eV rises from $0.47\times10^{-7}$ for Ag$_{255197}$ to $1.35\times10^{-7}$ for Ag$_{249}$ due to enhanced Landau damping of the surface plasmons.~\cite{Manjavacas2014,Govorov2017,Govorov2020} The HCs are effectively much hotter in smaller NPs. 

Figure~\ref{fig:fig2}(a) shows the size dependence of the total number of HCs (black dashed line) and the high-energy portion with excitation energies larger than 2~eV (black solid line). The latter is expected to be more efficient for vibrational coupling and bond dissociation. These quantities are obtained by integrating $\int \mathrm{d}\epsilon\,\delta\rho(\epsilon)$ in given energy windows.~\cite{Govorov2017,Lischner2023_1} The total number of HCs increases nonlinearly, exhibiting a $D^3$ scaling as the total number of conduction electrons in NPs. In contrast, high-energy HCs are generated by plasmon damping via the surface scattering,~\cite{Govorov2013,Manjavacas2014,Govorov2017,Govorov2020} which scales as $D^2$.~\cite{Govorov2017,Govorov2020} It can thus be expected that the rate of surface scattering exhibits a $1/D$ dependence. Therefore, smaller NPs generate a higher portion of energetic HCs. The $1/D$ scaling of Landau damping has been revealed in early studies using the electron gas model.~\cite{Govorov2017,Govorov2020} However, how this size dependence affects the vibrational excitation and catalysis remains unexplored.

\subsection{\label{sec:RD2} Size-dependent vibrational excitation and dissociation}
\begin{figure*}[h!tbp]
	\includegraphics[width=0.8\columnwidth]{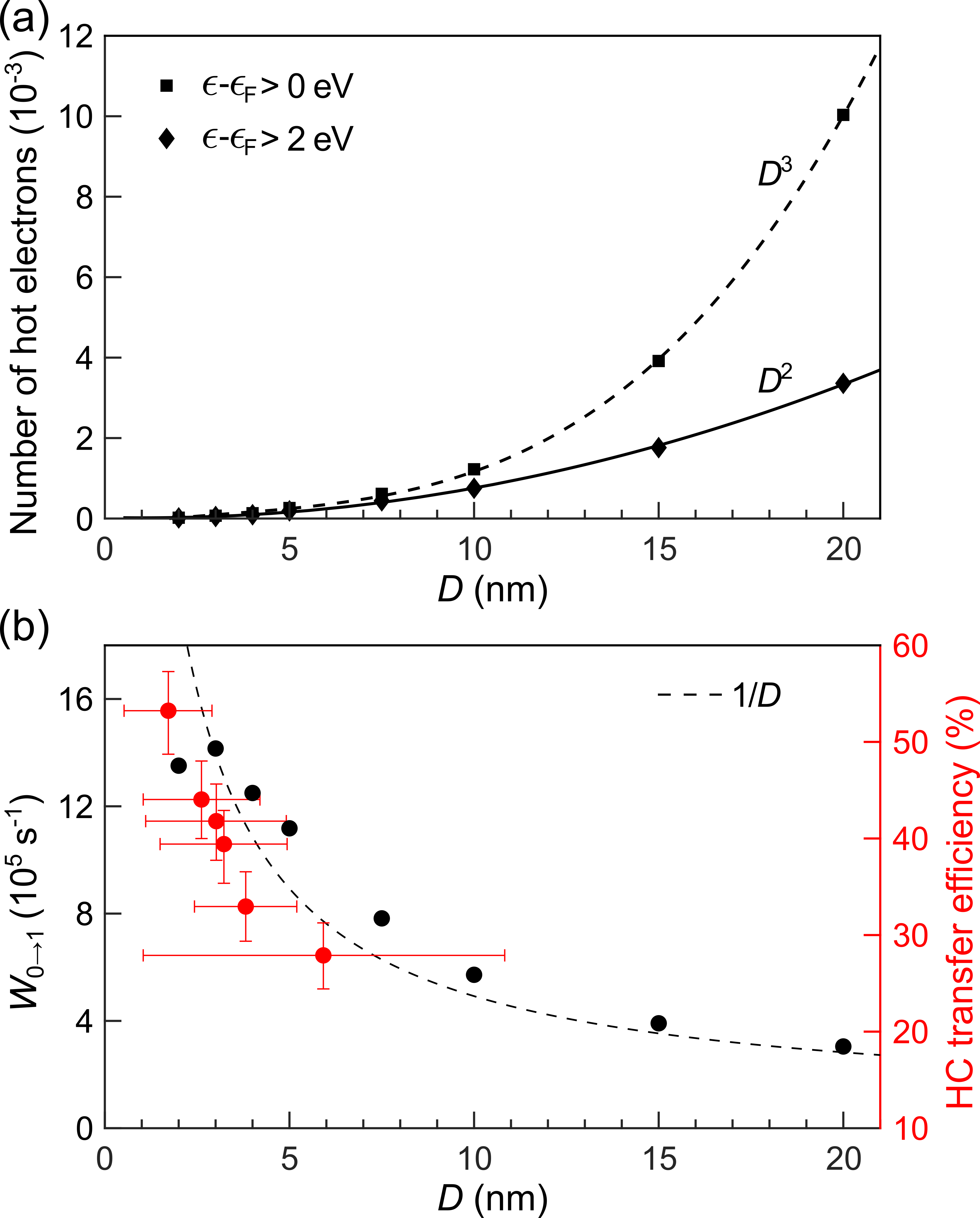}
	\caption{\label{fig:fig2} Size-dependent generation of HCs and their induced vibrational excitation rates. (a) The total number of HCs (squares) and the number of HCs with energies above 2~eV (diamonds) as a function of $D$. Spectral fitting reveals that they scale as $D^3$ (black dashed line) and $D^2$ (black solid line), respectively. (b) Size dependence of the vibrational excitation rates, $W_{0\to 1}$, of O$_2$ stretch mode on Ag NPs (black circles). The results exhibit good consistency with the HC transfer efficiency (red circles, digitized from Ref.~\cite{Lian2021}), both showing an overall 1/$D$ scaling (black dashed line). Molecular adsorption parameters are adopted from Ref.~\cite{O2_2012}. All calculations are performed with a light intensity of 10$^6$~W/m$^2$.} 
\end{figure*}

Figure~\ref{fig:fig2}(b) shows the rate of vibrational excitation $W_{0\to 1}$ for O--O stretch mode on Ag NPs (the black circles) as a function of $D$, which is calculated from Eq.~\eqref{eqn:vib_exc}. As $D$ decreases, $W_{0\to 1}$ exhibits a significant 5-fold increase due to enhanced vibrational coupling to HCs in the high-energy region. The model parameters for adsorption and electron-vibration coupling are extracted from O$_2$ on Ag(100).~\cite{O2_2011,O2_2012} Overall, the vibrational excitation rate scales as $1/D$, as indicated by the black dashed line in Fig.~\ref{fig:fig2}(b). This scaling results from the size-dependent plasmon damping of NPs,~\cite{Govorov2017,Govorov2020} as demonstrated in Fig.~\ref{fig:fig2}(a). Interestingly, it is also consistent with the size-dependent HC injection efficiency measured at Ag/TiO$_2$ interface~\cite{Lian2021} (cf. red dots in Fig.~\ref{fig:fig2}(b)). While the general trend holds for large and intermediate-sized NPs, deviations and oscillations emerge for $D\le 5.0$~nm due to the energy level quantization in smaller NPs. This oscillatory behavior becomes even more prominent with increased HC lifetimes (Fig.~S3), as can be expected due to the shrinking phase space for carrier relaxation.~\cite{Nth_exp_2025,Schatz2026,Abajo2016,Liu2017}

\begin{figure*}[h!tbp]
	\includegraphics[width=0.8\columnwidth]{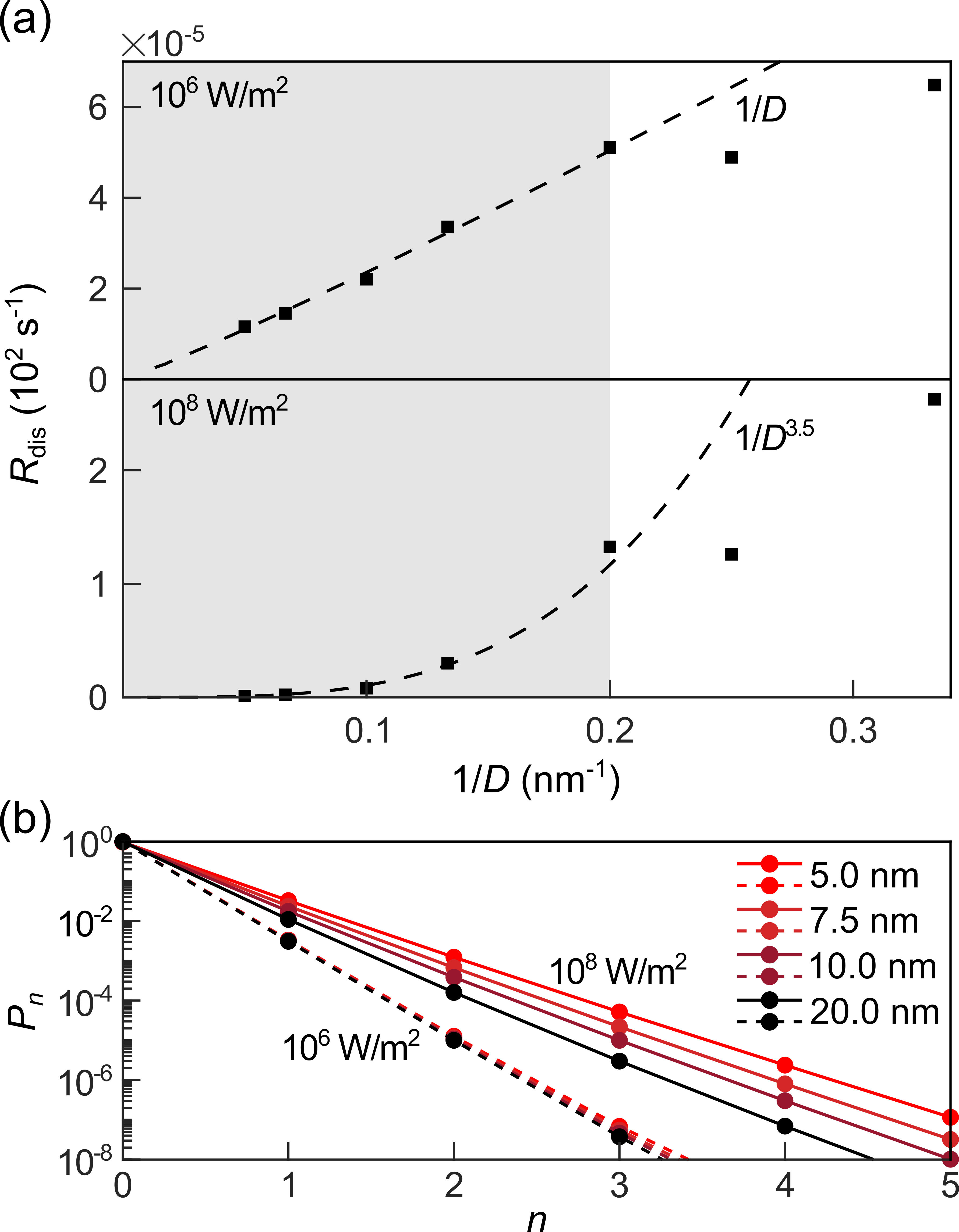}
	\caption{\label{fig:fig3} Vibrational activation and photodissociation induced by plasmonic HCs. (a) Size-dependent O$_2$ dissociation rates evaluated under different light intensities. The shaded area indicates the size regime with $D>5.0$~nm, in which the power-law fitting of the dissociation rate yields a linear $1/D$ scaling at 10$^6$~W/m$^2$, and a nonlinear $1/D^{3.5}$ scaling at 10$^8$~W/m$^2$. (b) The steady-state vibrational distribution $P_n$ of adsorbed O$_2$ on four representative Ag NPs. The dashed and solid curves correspond to the results obtained under different light intensities. All calculations are performed with $\epsilon_a$~=~2.4~eV and an ambient temperature of 200 K.}
\end{figure*}

The enhancement of vibrational excitation in smaller NPs has direct implications for the rate of photodissociation,~\cite{Kim2023,Kim2024} $R_\mathrm{dis}=\sum_n P_n W_{n\rightarrow n_\mathrm{d}}$. It involves the inelastic transition rates from both the ground state $n=0$ and the vibrationally excited states $n\ge1$ within the truncated harmonic oscillator approximation. In the low-light-intensity and low-temperature regimes, the molecule is essentially in the vibrationally ground state. The dissociation rate is therefore dominated by single inelastic transition from the ground state to the lowest dissociated state $n_\mathrm{d}$, namely $R_\mathrm{dis}\sim W_{0\to n_\mathrm{d}}$. It suggests a $1/D$ dependence of the dissociation rate within this regime. The upper panel of Fig~\ref{fig:fig3}(a) shows the calculated dissociation rate as a function of $1/D$ at a low light intensity of 10$^6$~W/m$^2$. Indeed, a linear scaling is found as expected. In this case, the vibrational population remains the same for all $D$, which follows the Boltzmann distribution at $T=200$~K as shown in Fig.~\ref{fig:fig3}(b).

At higher light intensities, local vibrational heating occurs in the molecular bond due to multiple inelastic scatterings.~\cite{Gao1997,Olsen2009,Newns1992} The dissociation rate has additional and even dominant contributions from the vibrationally excited states. This may lead to a different size dependence in photocatalytic reactions that deviates significantly from $1/D$. The lower panel of Fig~\ref{fig:fig3}(a) shows the dissociation rates at a higher intensity of 10$^8$~W/m$^2$, where the populations of vibrationally excited states are significantly enhanced in smaller $D$, as shown in Fig~\ref{fig:fig3}(b). The power-law fitting of the dissociation rate for $D=5.0-20.0$~nm suggests a nonlinear and stronger 1/$D^{3.5}$ scaling, with a more pronounced rate enhancement as $D$ goes down toward the quantum-size regime. For instance, $P_4$ has increased by a factor of 34.2 as $D$ goes from 20.0~nm to 5.0~nm. The corresponding dissociation rates are enhanced by a factor of 110.1 at 10$^8$~W/m$^2$ compared to 4.4 at 10$^6$~W/m$^2$. As a result, HCs in smaller NPs can accelerate catalytic reactions much more effectively in the vibrational heating mechanism. 

\subsection{\label{sec:RD3} Selective vibrational coupling by quantized HC spectra}
\begin{figure*}[h!tbp]
	\includegraphics[width=0.8\columnwidth]{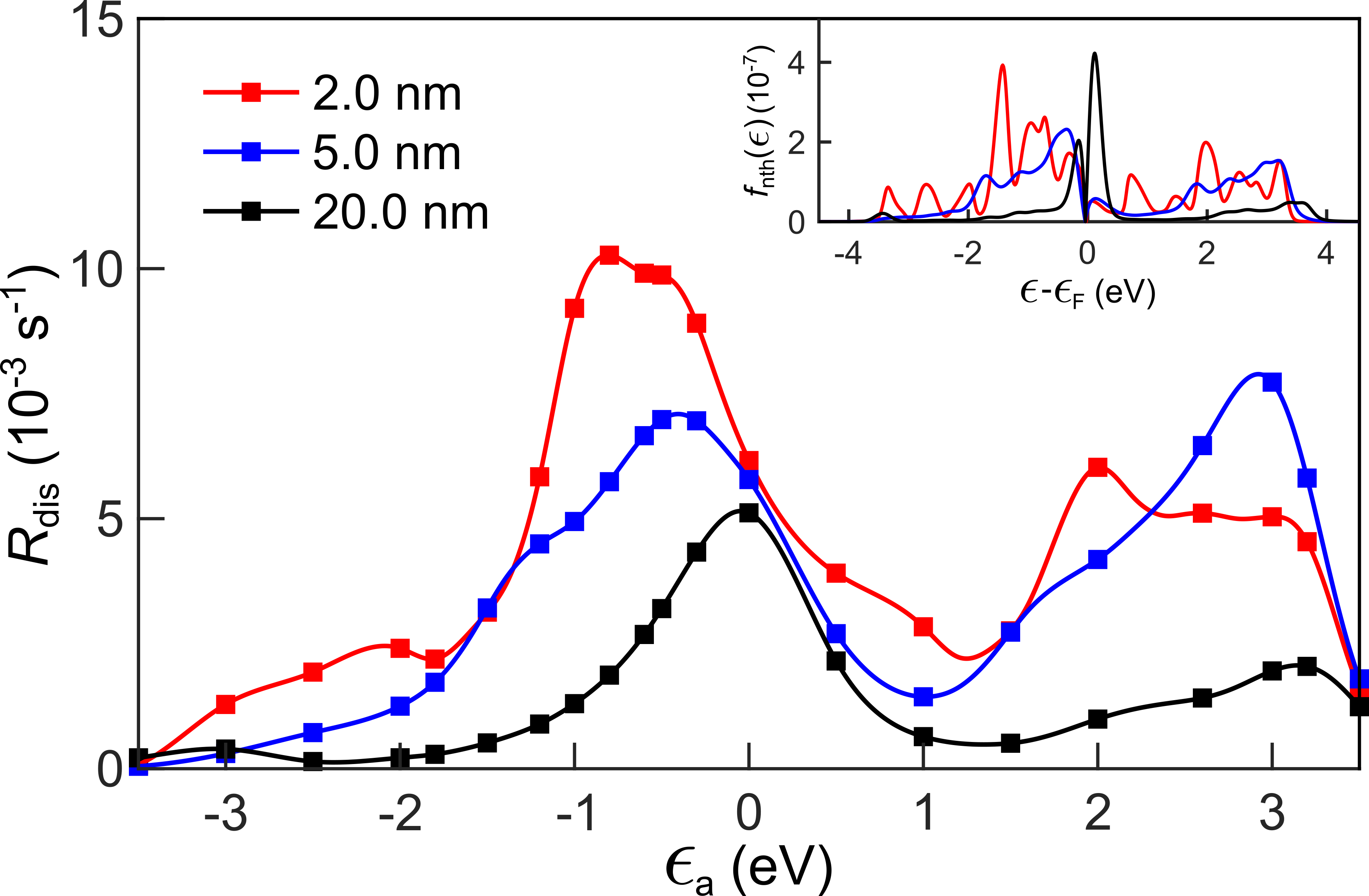}
	\caption{\label{fig:fig4} Resonance energy dependence of O$_2$ dissociation rate on three representative NPs, with $D=2.0$~nm, 5.0~nm, and 20.0~nm. The inset shows their corresponding HC spectra. Light intensity and temperature employed in calculations are 10$^6$~W/m$^2$ and 200~K.}
\end{figure*}

The quantized HC spectra in smaller NPs are also promising for selective control of vibrational excitation and catalytic reactions. Adsorbed molecules can form both occupied and unoccupied molecular resonances due to coupling with metal electrons.~\cite{Nordlander2024} These molecular resonances, as captured by the parameter $\epsilon_a$, have maximal coupling to those HCs with excitation energies in the same energy region of $\epsilon_a$.~\cite{NH3_2022,NH3_2024,Christopher2014,Alabastri2024} Here, negative $\epsilon_a$ corresponds to hole resonance, which could be formed by hole scattering.~\cite{Palmer1994} Figure~\ref{fig:fig4} shows the dissociation rates as a function of resonance energy $\epsilon_a$ for $D=20.0$~nm, 5.0~nm, and 2.0~nm, respectively. The results exhibit two general peaks, one at 2--3 eV and another spanning over -1~eV to 0~eV energy ranges. These two peaks arise from the resonance coupling to hot electron and hot hole peaks, respectively, and follow the general spectral features of the HC distributions as shown in the inset. Reducing the particle sizes not only boosts the reaction rates, but also modulates the energy positions of the maximum rates. As $D$ decreases from 20.0~nm to 2.0~nm, the maximum of the high-energy peak shifts from $\epsilon_a=3.2$~eV to 2.0~eV, while the low-energy peak shifts from $\epsilon_a=0.0$~eV to -0.8~eV. Obviously, the highest dissociation rate peaks vary with particle sizes. In large NPs with $D=20.0$~nm, thermal carriers distributed near $\epsilon_\mathrm{F}$ contribute dominantly to the reaction through the resonance coupling to $\epsilon_a=0.0$~eV. In contrast, the mechanism and rate of dissociation in smaller NPs are dominated by different carriers, namely hot electrons coupled to $\epsilon_a=2.9$~eV for $D=5.0$~nm, and hot holes coupled to $\epsilon_a=-0.8$~eV for $D=2.0$~nm, respectively. This suggests that quantum oscillations in the HC spectra can alter the coupling mechanism, electron vs. hole, in catalytic reactions. It offers promising route to size-selective photocatalysis using the quantized HC spectra in different plasmonic structures.

The above results were obtained by a set of parameters for optical absorption~\cite{CRCbook} and molecular coupling derived from O$_2$ on bulk Ag surfaces, and did not take into account of possible size-dependent changes such as the plasmon frequencies~\cite{PR2012,PR2018,PR2024} and HC lifetimes in finite NPs.~\cite{Link1999,Broyer2000,Treguer2003} We have carrier out simulations in a wide range of parameters of the model, and DFT calculations for O$_2$ adsorbed on small NPs (Fig.~S4). For example, with a narrower resonance width $\Delta_a$, which is expected in smaller NPs, the dissociation rates become much higher than those in Fig.~\ref{fig:fig4} (cf. Fig.~S5). The maximal reaction peaks also shift more sensitively with molecular resonances in smaller NPs. While the quantitative details of results may change with these parameters, the main conclusions remain essentially the same. As another remark, our model adopts a global energy distribution without discriminating the surface states. With local energy distribution, which might differ slightly from its global counterpart,~\cite{Erhart2020} we expect that the increased surface sensitivity would strengthen our basic conclusion. 

\section{\label{sec:Conclusion} Conclusion}
In conclusion, we have investigated the HC-induced vibrational excitation and dissociation of O$_2$ on Ag NPs with diameters spanning from the classical to quantum size regime. Our results reveal a $1/D$ dependence in the vibrational excitation rate, which can be attributed to the size-dependent Landau damping of surface plasmons. Pronounced oscillations emerge as $D$ approaches the quantum-size regime due to electron energy quantization. Vibrational heating at high light intensities can effectively pump the vibrational populations into higher excited states, leading to nonlinear size dependence and stronger enhancement in the reaction rate. The emergence of quantized HC distributions in smaller NPs not only boosts the efficiency of the photocatalytic reaction, but also offers a promising route for selective molecular coupling and the plasmon-driven reactions. Our findings call for research in plasmonic catalysis in the quantum size regime. 

\section*{Supplementary Material}
See the supplementary material for (1) the implementation details of KPM; (2) the evolution of the density of with the NP size; (3) size-dependent vibrational excitation rates under varied HC lifetimes; (4) DFT calculations for O$_2$ adsorbed on small Ag clusters; and (5) the resonance energy dependence of molecular dissociation rate with a narrower resonance width.

\section*{Acknowledgments}
This work was supported by Science Challenge Project (Grant No. TZ2025013), National Natural Science Foundation of China (Grant Nos. 12393831 and 11934003). J. Lischner acknowledges funding from the Leverhulme Trust (RPG-2024-317) and the EPSRC program grant EP/W017075/1. Computational resources are provided by Tianhe2-JK at CSRC. We thank Prof. Lishi Luo for critical reading on the manuscript. 

\section*{Author Declarations}
\subsection*{Conflict of Interest}
The authors have no conflicts to disclose.

\subsection*{Author Contributions}
\textbf{Yu Chen:} Data curation (lead); Formal Analysis (equal); Methodology (equal); Software (equal); Writing--original draft (lead); Writing--review and editing (equal). \textbf{Hanwen Jin:} Formal Analysis(equal); Methodology (equal); Software (equal); Writing--review and editing (equal). \textbf{Fei Gao:} Software (equal); Writing--review and editing (equal). \textbf{Johannes Lischner:} Formal Analysis (equal); Methodology (equal); Software (equal); Writing--review and editing (equal). \textbf{Shiwu Gao:} Conceptualization (lead); Formal Analysis (equal); Methodology (equal); Writing--original draft (supporting); Writing--review and editing (equal).

\section*{Data Availability}
The data that support the findings of this study are available from the corresponding author upon reasonable request.

\clearpage

\clearpage

\clearpage

\clearpage

\clearpage

\bibliography{ref_jcp}

\clearpage
\section*{Supplementary Material for: Quantum hot carrier spectra in plasmonic catalysis}
\setcounter{section}{0}
\setcounter{figure}{0}
\setcounter{table}{0}
\setcounter{equation}{0}
\renewcommand{\thesection}{S\arabic{section}}
\renewcommand{\thefigure}{S\arabic{figure}}
\renewcommand{\thetable}{S\arabic{table}}
\renewcommand{\theequation}{S\arabic{equation}}

\section{Numerical details of the theoretical framework}
\noindent\textbf{Model Hamiltonian} Our theory~\cite{Wu2022,Gao2025} is based on the Anderson-Newns model~\cite{Newns1969} for electron-molecule scatting.~\cite{Gadzuk1991,Gao1997,Olsen2009} The Hamiltonian of the combined nanoparticle(NP)-molecule system is given by
\begin{equation}\label{eqn:H_AN}
	\begin{split}
		H&=\sum_k \epsilon_k c_k^\dagger c_k+\epsilon_a c_a^\dagger c_a+\sum_k (v_{ak} c_a^\dagger c_k+\mathrm{H.c.})\\
		&\hspace{2em}+\hbar\Omega b^\dagger b+\lambda c_a^\dagger c_a(b^\dagger+b).
	\end{split}
\end{equation}
Here, the first line describes the tunneling coupling between an electron state $\{\epsilon_k; |k\rangle\}$ of the NP and a molecular resonance $|a\rangle$ of energy $\epsilon_a$, with the coupling $v_{ak}$. The molecular bond is simulated by a truncated harmonic oscillator of frequency $\Omega$, which is linearly coupled to the electron bath with a strength $\lambda$.

Within the resonance electron-molecule scattering formalism,~\cite{Gadzuk1991} inelastic vibrational transition rate $W_{n\to n^\prime}$ between vibrational states $n$ and $n^\prime$ was evaluated as Eq.~(3) of the main text.~\cite{Gao1997} It requires two sets of inputs: (1) the molecular resonance parameters $\epsilon_a$ and $\Delta_a$, which incorporate all tunneling coupling effects in the wide-band approximation,~\cite{Gadzuk1991,Gao1997,Olsen2009} and (2) the electronic distribution function $f(\epsilon)$ of the electron bath of the NPs. For the ground state, $f(\epsilon)=f_0(\epsilon)$, where $f_0(\epsilon)$ is the Fermi-Dirac distribution at a given temperature. Under plasmon excitation, a nonthermal distribution function $f_\mathrm{nth}(\epsilon)$~\cite{Govorov2020} was introduced to account for the contributions from nonequilibrium hot carriers (HCs) generated by plasmonic damping.~\cite{Wu2022,Gao2025}

\bigskip
\noindent\textbf{Tight-binding description of the NPs} To compute the density response and the nonthermal distribution $f_\mathrm{nth}(\epsilon)$, single-particle eigenspectrum $\{\epsilon_k; |k\rangle\}$ of the NP is needed. It requires diagonalzation of the NP Hamiltonian $H_\mathrm{NP}$, which is constructed in the tight-binding approximation. Atomic structures of nearly spherical Ag NPs are constructed from the face-centered-cubic lattice of the bulk crystal. Specifically, we first select one atom as the particle center and then retain all atoms located within given radius from this center atom.~\cite{Lischner2022,Lischner2023_1} Taking the valence orbitals of Ag (4$d$, 5$s$ and 5$p$) as a basis set, the tight-binding Hamiltonians $H_\mathrm{NP}$ of NPs are constructed through an orthogonal two-center Slater-Koster parameterization.~\cite{Papaconstantopoulos2015} 

\bigskip
\noindent\textbf{KPM for the density of states} Having scaled the energy variable ($\epsilon\to\varepsilon$) and Hamiltonian spectrum ($H_\mathrm{NP}\to \tilde{H}$) to $[-1,1]$ for appropriate definitions of the first-kind Chebyshev polynomials $T_n$, the density of states (DOS) $g(\epsilon)$ (Eq.~(1) of the main text) can then be evaluated by expanding the spectral operator in terms of $T_n$ as~\cite{Sankey1993,Silver1994,Wang1994}
\begin{equation}\label{eqn:DOS_KPM_expansion}
	g(\epsilon)=\frac{2}{B\pi\sqrt{1-\varepsilon^2}}\sum_{n=0}^{N-1}\frac{T_n(\varepsilon)J(n, N)}{1+\delta_{n0}}\mathrm{Tr}[T_n(\tilde{H})].
\end{equation} 
Here, $B$ is the spectral scaling factor determined by the upper and lower bounds of the Hamiltonian eigenspectrum,~\cite{RMP2006} $J(n,N)$ is the Jackson's kernel employed to suppress the Gibbs oscillations caused by truncating the polynomials at finite order $N$. For the evaluation of $\mathrm{Tr}[T_n(\tilde{H})]$,  a linear-scaling stochastic trace evaluation technique~\cite{Sankey1993,Silver1994,Wang1994} was applied, 
\begin{equation}\label{eqn:STE}
	\mathrm{Tr}[T_n(\tilde{H})]=\frac{1}{R}\sum_{r=0}^{R-1}\langle r|T_n(\tilde{H})|r\rangle.
\end{equation}
Here, $|r\rangle=(\xi_r^1,\xi_r^2,\dots,\xi_r^K)^\mathrm{T}$ is a random vector with the same dimension as $H_\mathrm{NP}$, whose components $\xi_r^j$ are stochastic variables following the uniform distribution over $[-\sqrt{3},\sqrt{3}]$. We used $N=500$ and $R=200$ to ensure a converged $g(\epsilon)$, and took a Gaussian convolution with 0.1~eV width to obtain smooth spectra. 

\begin{figure*}[htbp]
	\includegraphics[width=\columnwidth]{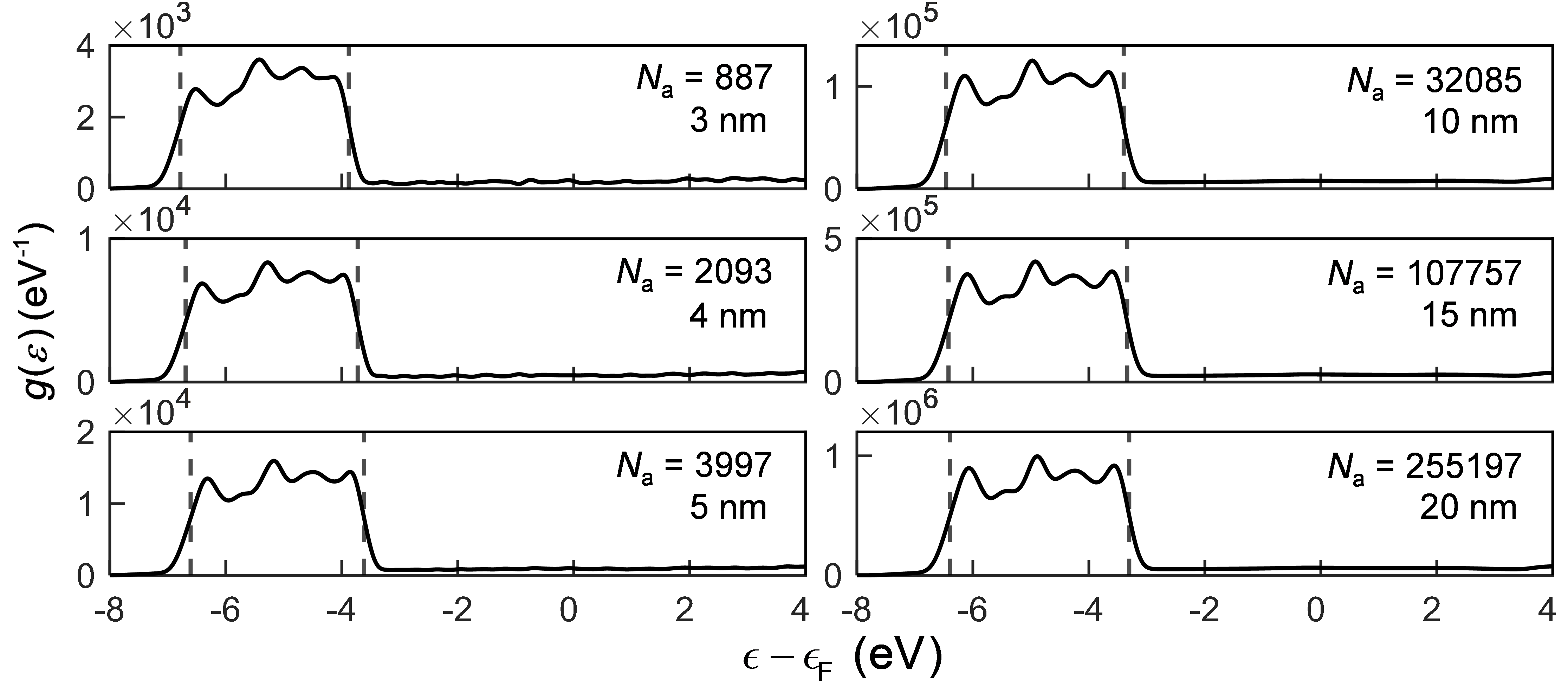}
	\caption{\label{fig:figS1} Electronic DOS for Ag NPs with different sizes obtained from the tight-binding model. Dashed lines indicate the energies at which the $d$-band DOS reaches half of its maximum value.}
\end{figure*}

\begin{figure*}[!htbp]
	\includegraphics[width=0.8\columnwidth]{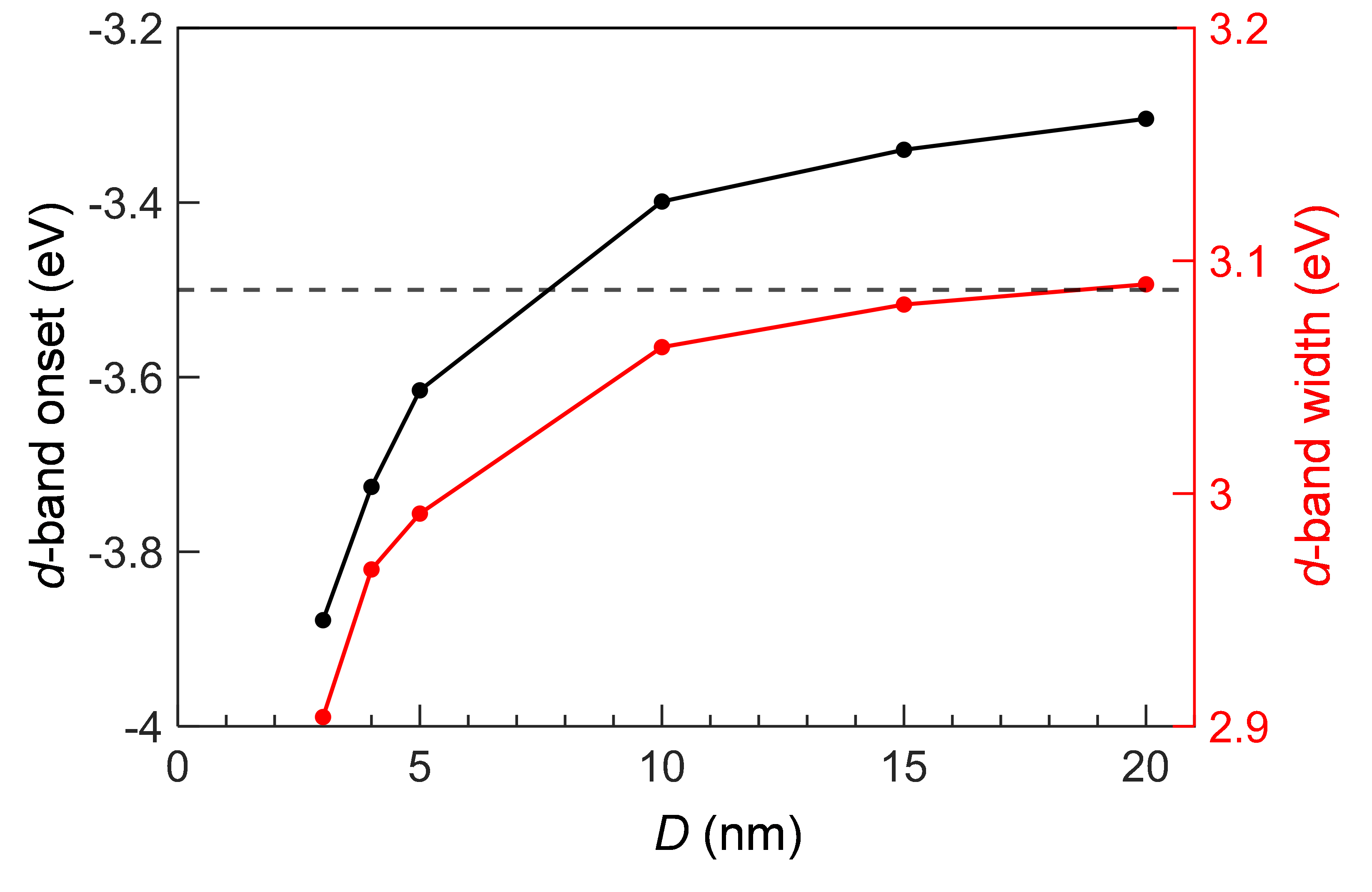}
	\caption{\label{fig:figS2} Size-dependent shifts of $d$-band onsite positions (left axis, black dots), and changes of $d$-band widths (right axis, red dots). The effective energy range of $d$-band is defined by the dashed lines shown in Fig.~\ref{fig:figS1}.}
\end{figure*}

Figure~\ref{fig:figS1} shows the results for six different sizes, with the largest one reaching $D=20.0$~nm, containing as many as $N_a=2\times 10^6$ Ag atoms. It is found that $g(\epsilon)$ exhibits a quasi-continuous distribution for $D\geq 10.0$~nm, and becomes increasingly discrete when the size is reduced to $D<5.0$~nm. Simultaneously, the $d$-band edge shifts downward while the $d$-band width narrows in smaller NPs due to the reduced electronic interaction and quantized electronic spectra, as illustrated in Fig.~\ref{fig:figS2}.

\bigskip
\noindent\textbf{KPM for the HC energy distribution} We followed the quantum master equation formulation developed by Govorov and coworkers to model the energy distribution of plasmonic HCs.~\cite{Govorov2013,Govorov2017} It has been recast in the KPM-compatible form as Eq.~(2) of the main text, with the optical matrix elements $\Phi_{kk^\prime}(\omega)$ between two discrete electronic states being transformed into a spectral density and can be treated in KPM as~\cite{Lischner2022,Lischner2023_1}
\begin{equation}\label{eqn:phi_KPM_expansion}
	\begin{split}
		\phi(\mathcal{E},\mathcal{E}^\prime,\omega)
		&=\mathrm{Tr}[\delta(\mathcal{E}-H_\mathrm{NP})\Phi^\dagger(\omega)\delta(\mathcal{E}^\prime-H_\mathrm{NP})\Phi(\omega)]\\
		&=\frac{4}{B^2\pi^2\sqrt{(1-\varepsilon^2)(1-\varepsilon^{\prime 2})}}\sum_{m,n=0}^{N-1}\frac{T_m(\varepsilon)J(m, N)}{1+\delta_{m0}}\\
		&\hspace{2em}\times\frac{T_n(\varepsilon^{\prime})J(n,N)}{1+\delta_{n0}}\mathrm{Tr}[T_m(\tilde{H})\Phi^\dagger(\omega)T_n(\tilde{H})\Phi(\omega)].
	\end{split}
\end{equation}
Analogous to the KPM evaluation of $g(\epsilon)$, the last line employs the spectral scaling and the Chebyshev expansion of the spectra operators, where the trace can be efficiently computed using the stochastic trace evaluation technique as Eq.~\eqref{eqn:STE}. 

During the numerical implementation, the optical field was treated within the quasistatic approximation,~\cite{Lischner2022,Lischner2023_1} and the corresponding optical matrix elements were expanded in the tight-binding basis.~\cite{Pedersen2001} To ensure convergent HC spectra for different particle sizes, the numbers of Chebyshev polynomials and random vectors were carefully chosen within the range of $N=2000-5000$ and $R=2000-8000$.

\newpage

\clearpage

\section{The effect of HC lifetime on the vibrational excitation rate}
\begin{figure*}[!htbp]
	\includegraphics[width=0.8\columnwidth]{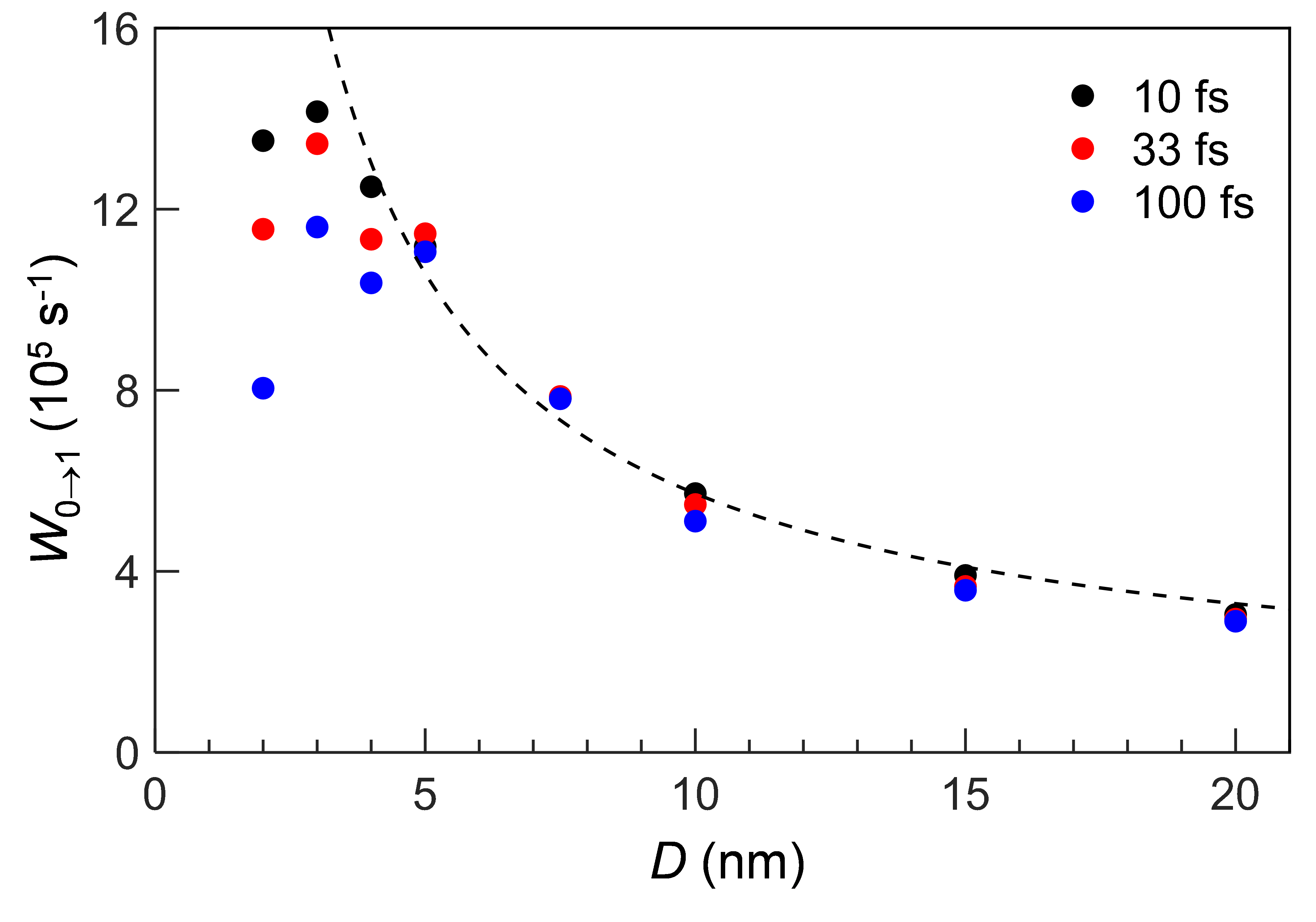}
	\caption{\label{fig:figS3} Size-dependent vibrational excitation rates, $W_{0\to1}$, evaluated in the atomistic model under different HC lifetimes $\tau_p$. Other parameters employed in the calculations are consistent with those in Fig.~2(b) of the main text.}
\end{figure*}

In the main text, the energy and momentum lifetimes of plasmonic HCs were set to $\tau_e=500$~fs and $\tau_p=33$~fs, respectively, following earlier work.~\cite{Govorov2017} Figure~\ref{fig:figS3} examines how the value of $\tau_p$ affects the size-dependent vibrational excitation rate $W_{0\to 1}$, as this parameter modulates the HC spectra, and is expected to vary in smaller NPs. The results show that, for $D\ge5.0$~nm, the $1/D$ scaling of $W_{0\to 1}$ remains consistent across different $\tau_p$. Nevertheless, the oscillatory behavior of $W_{0\to 1}$ becomes more prominent with larger $\tau_p$ in the quantum-size regime ($D<5.0$~nm).

\bigskip

\clearpage

\section{DFT calculations for molecular adsorption on silver clusters}
To understand how the electronic properties of adsorbed molecules vary with NP size, we performed DFT calculations for O$_2$ adsorbed on icosahedral Ag$_{55}$ and Ag$_{147}$ clusters~\cite{PR2024}, and compared the results with those for O$_2$/Ag(100). Structural optimization and spin-polarized electronic structure calculations were carrier out using ab initio plane-wave package VASP.~\cite{VASP1993} The projected-augmented wave (PAW)~\cite{PAW1,PAW2} and the general gradient approximation (GGA)~\cite{GGA1991} in PBE form for exchange-correlation energy~\cite{GGA1996} were used. We also applied a mean-field Hubbard correction of U = 3 eV to the O 2$p$ orbitals.~\cite{Lorente2010} The K-point mesh of 1$\times$1$\times$1 was used and all atoms were allowed to relax until the force on each atom had magnitudes less than 0.04 eV/\AA.

\begin{figure*}[htbp]
	\includegraphics[width=\columnwidth]{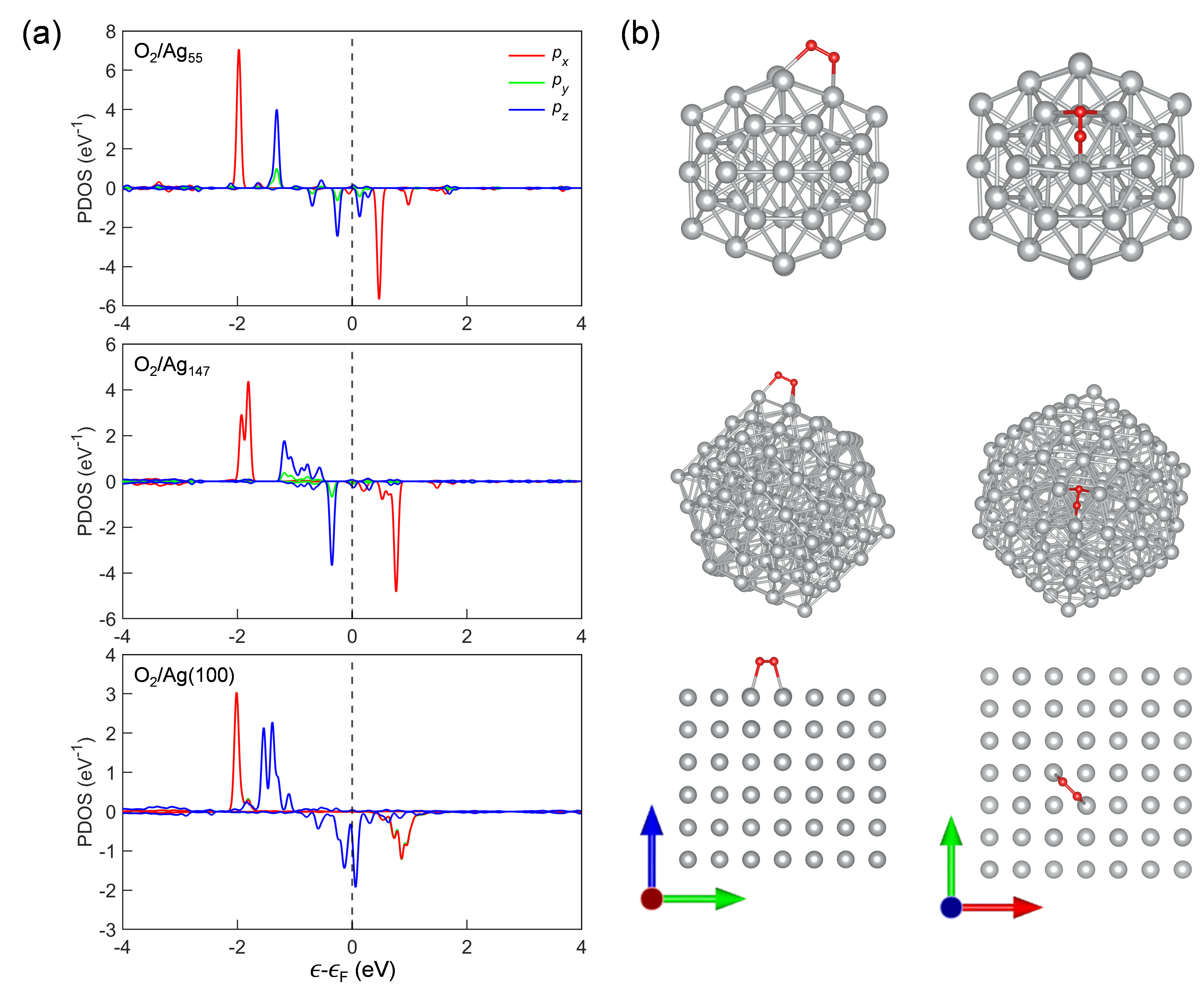}
	\caption{\label{fig:figS4} (a) The DOS projected on molecular $p$ orbitals for O$_2$ on Ag$_{55}$, Ag$_{147}$, and Ag(100). (b) Their corresponding stable adsorption configurations viewed along $x$ and $z$ directions. $x$, $y$ and $z$ directions are colored by red, blue, and green, respectively.}
\end{figure*}

Figure~\ref{fig:figS4}(a) shows the atomic-orbital-projected DOS (PDOS), with the optimized adsorption configures presented in panel (b). Compared with the adsorption on bulk surface, O$_2$ adsorbed on small Ag clusters exhibits significantly reduced broadening of the molecular PDOS. In particular, the $p_z$ PDOS of O$_2$/Ag(100) overlaps with the Fermi level, which splits into several distinct peaks and upon adsorption on small Ag clusters. Since this atomic orbital contributes to the partially occupied $2\pi^\ast$ molecular orbital of O$_2$ adsorbed on metal surfaces, the above findings indicate size-dependent variations in resonance energies and widths on NP surfaces.

\clearpage

\section{Resonance energy dependence of molecular dissociation rate}
\begin{figure*}[!htbp]
	\includegraphics[width=0.8\columnwidth]{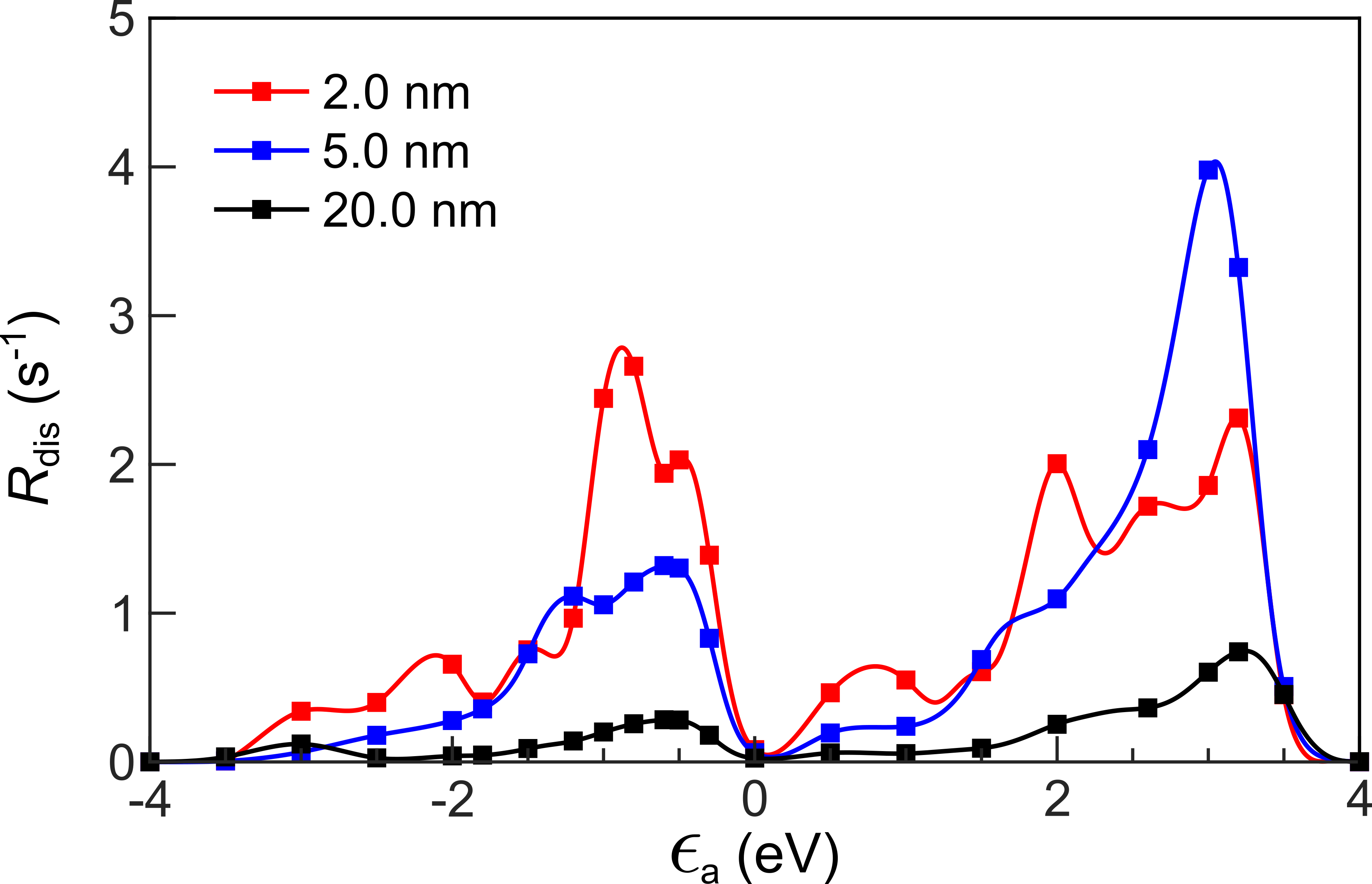}
	\caption{\label{fig:figS5} The resonance energy dependence of O$_2$ dissociation rates on three Ag NPs evaluated with $\Delta_a=0.3$~eV. Other calculation parameters are the same as Fig.~4 of the main text.}
\end{figure*}

A resonance width of $\Delta_a=0.6$~eV derived from O$_2$ adsorbed on Ag(100)~\cite{O2_2012} was used in the main text. This parameter correlates with the width of molecular PDOS, which can be reduced on smaller NPs as indicated by our DFT calculations. To account for this effect, Fig.~\ref{fig:figS5} evaluates the resonance energy dependence of molecular dissociation rate, following the analysis in Fig.~4 of the main text but using a smaller $\Delta_a=0.3$~eV. The reduction of resonance width corresponds to longer tunneling coupling lifetime, which benefits the nonthermal excitation pathway. As a result, the overall dissociation rates obtained here are much large than those in Fig.~4 of the main text. Besides, the narrower resonance improves the energy selectivity of its coupling to the quantized HC spectra in small NPs, giving rise to more pronounced shifts of maximal reaction peaks across different resonance energies.

\bigskip

\end{document}